\documentstyle[osa,manuscript]{revtex}  
\begin{document}                
\title{Connection between topology and statistics}

\author{Zhong Tang *  and  David  Finkelstein **}

\address{School of Physics, Georgia Institute of 
Technology, Atlanta, Georgia 30332-0430}

\maketitle
\begin{abstract}
 
We introduce topological magnetic field in two-dimensional 
flat space, which admits a solution of scalar monopole
that describes the nontrivial topology. 
In the Chern-Simons gauge field theory of anyons, 
we interpret the anyons as the quasi-particles composed of  
fermions and scalar monopoles in such a form that each fermion 
is surrounded by infinite number of scalar monopoles.
It is the monopole charge that determines the statistics 
of anyons. We re-analyze the conventional arguments
of the connection between topology and statistics 
in three-dimensional space, and find that those arguments are 
based on the global topology, which is relatively trivial
compared with the monopole structure. Through a 
simple model, we formulate the three-dimensional anyon field 
 using infinite number of Dirac's magnetic monopoles 
that change the ordinary spacetime topology. 
The quasi-particle picture of the three-dimensional anyons is
quite similar to that of the usual two-dimensional anyons.
However, the exotic statistics there is not 
restricted to the usual fractional statistics,
but a functional statistics.\\

\noindent PACS number(s): 05.30.-d, 03.65.-w, 02.40.+m.\\

\noindent 
Electronic addresses: 

 *gt8822b@prism.gatech.edu, 
 **david.finkelstein@physics.gatech.edu.

\end{abstract}

\newpage
\begin{center}
\section*{I.~ Introduction}               
\end{center}

More than fifty years ago,  based on local relativistic
quantum field theory, Pauli \cite{pauli} proved  
that spin-integral particles  obey bose statistics, 
and spin-half integral particles 
obey fermi statistics. Since then, the connection
between spin and statistics has become an interesting 
topic and acquired a lot of studies. 
Among the earlier attempts to study  physics
using topology, Ref. [2] is probably the first
to adopt homotopy group to describe the intrinsic
invariant for example  $kink$ in some nonlinear systems
embedding soliton.
Through an analysis of the significance of 
topology for quantization of nonlinear classical fields, 
one of the authors and Rubinstein \cite{finkelstein} clarified 
that in the Skyrme's model \cite{skyme}, 
the half-integral spin solitons  obey fermi statistics,
and showed that statistics is related with topology. 
Leinaas and Myrheim,\cite{leinaas} and later Wilczek 
\cite{wilczek} demonstrated that in two spatial dimensions, 
the possibilities for quantum statistics are not limited to 
bosons and fermions, but rather allow continuous interpolation
between these two extremes, called exotic statistics, which 
is defined by the phase of the amplitude associated with slow 
motion of distance particles around one another. 
The particles following the exotic statistics are
generally called $anyons$.
The discovery of anyons in two spatial 
dimensions opens a new ground for the study
in condensed matter system. 
The manifestations of anyons as quasi-particles 
in fractional quantum Hall effect \cite{laugh} and anyon 
superconductivity \cite{chen} have been intensively 
investigated for years. 

Since the two-dimensional 
Abelian rotation group allows fractal angular momentum,  
in quantum mechanics 
exotic statistics is interpreted through the 
trajectories of the wavefunction of the objects with 
fractal angular momentum \cite{leinaas,wilczek}.
Nevertheless, there are at least three spatial 
dimensions in the physical world, it is clear that anyons
are not real particles. In quantum field theory, 
the anyons are interpreted as the quasi-particles through a 
system of $spinless$ fermionic field which minimally
couples to the Chern-Simons gauge field \cite{fradkin}. 
It is the Chern-Simons term, a topological invariant
of winding number one, that transforms the fermions 
into anyons. Recently, the Chern-Simons gauge theory was 
found to embed the solution of solitons, 
which still follow an exotic statistics \cite{frohlich}. 
There is a further question:

How to interpret the exotic statistics 
from the viewpoint of topology \cite{comment1}? 

In the Chern-Simons gauge theory of anyons, 
the above question becomes more concrete:
What kind of topology does the Chern-Simons theory describe 
in the system of anyons? In this paper we attempt to answer 
this question by studying a system of nonrelativistic 
fermionic field, which minimally couples to the Chern-Simons 
gauge field. Through looking into the two-dimensional 
topology, we find that the system embeds a solution of 
topological monopole, called $scalar ~monopole$ (its
potential is a scalar), which describes the two-dimensional 
nontrivial topology  analogous to the magnetic monopole of 
three-dimensional space. As a result, the anyons are 
found to be such combinations of fermions and scalar monopoles
that each fermion is surrounded by infinite number of scalar 
monopoles in the whole space. Within this quasi-particle picture,
the exotic statistics is easily interpreted:  
When two anyons exchange their positions, 
each of them must pass through the scalar monopole
of the others, and result in a nontrivial 
phase factor in term of the $charge$ of the scalar monopole
that determines the statistics of the anyons.
 
Another topic discussed in the present paper is about the 
statistics in three-dimensional space.
It has been argued \cite{leinaas} from the 
viewpoints of both topology and quantum field theory
that there are only bose and fermi statistics in three spatial 
dimensions. Nevertheless, there are still some interesting
discoveries worth notice: Both the Dirac's magnetic monopole and 
t'Hooft-Polyakov's non-Abelian monopole \cite{polyakov} are 
described by the bosonic fields, however, they obey fermi 
statistics \cite{monosta}; The bosonization scheme was
realized in four-dimensional quantum electrodynamics 
\cite{boseqed}
using the Wilson loop;
The Jordan-Wigner transformation was generalized to three
spatial dimensions by $SU(2)$ doublet Heisenberg spin 
operators \cite{huerta}, etc. We notice that though these
discoveries enrich the three-dimensional statistics, 
they are still consistent with the usual topological 
argument of statistics. 
But, there is one exception as far as we know: 
Libby, Zou and Laughlin \cite{libby} formulated fractional 
statistics in three-dimensional chiral spin 
liquid state through $SU(2)$ 
non-Abelian gauge interaction. Unfortunately, 
the fractional statistics there, defined by Berry's phase,
breaks the three-dimensional rotational invariance. 

In this paper, we re-analyze the conventional arguments of the 
three-dimensional statistics, and find that the usual conclusion
results from the relatively trivial topology,
in other words, there is no nontrivial topological structure 
like the scalar monopole of two dimensions
involved in the system as the background topology.
Then, are there other possibilities of statistics besides 
bosons and fermions in three-dimensional space with $nontrivial$ 
topology? To answer this question, we consider a system of
non-relativistic fermion field, which couples to an Abelian 
gauge field. The dynamics of the gauge field is 
determined by the Pontrjagin term multiplying a scalar field. 
By solving the Lagrangian equations, we find that the 
gauge field can be formally expressed as a pure gauge in 
terms of the vector potential of Dirac's magnetic monopole. 
Then through a peculiar gauge transformation, we construct
an $anyon$ field by fermions and magnetic monopoles, 
where each fermion is surrounded by 
infinite number of magnetic monopoles in the whole space.
This quasi-particle picture of anyons is similar to that of 
the usual two-dimensional anyons. However, the exotic 
statistics is not restricted to be the fractional 
statistics, but becomes functional statistics. 
Same as the case in Ref. [17], the anyons in our model  
break the rotational invariance too, which implies that 
the three-dimensional anyons can only be realized as 
the quasi-particles in condensed matter system, where the
rotation symmetry is not necessarily preserved in many cases.  
We further point out that the action of the anyons
results from a low-energy approximation for a  quantum system 
with axial chiral current anomaly. The scalar field is then 
consistently interpreted as the chiral mode of the system.

This paper is organized as follows: In Section {\bf II}, 
we first introduce topological magnetic field and 
scalar monopole in two-dimensional space, then 
show how the scalar monopole enters the usual Chern-Simons
gauge field and determines the statistics of anyons. In addition,
we point out that the Chern-Simons field can be 
realized by Berry's potential.  
Section {\bf III} is devoted to the connection between 
statistics and three-dimensional topology. We present
a simple model to formulate
the exotic statistics in three-dimensional space. 
Section {\bf IV} is the conclusion.\\

\begin{center}
\section*{II. ~Topology and statistics in (2+1)-dimensional 
spacetime}
\end{center}

As we know in  quantum mechanics, 
the exotic statistics in two-dimensional space 
is interpreted by the objects with fractal angular
momenta; On the other hand, in the Chern-Simons gauge 
field theory, it is interpreted by the gauge flux
produced by the permutation of particles.
In this section, we provide an alternative 
interpretation for the exotic statistics from the
viewpoint of topology.
Usually, one studies the topology through the physical models. 
For an example, topological defect in three-dimensional space 
is modeled by Dirac's magnetic monopole.
In the following Subsection {\bf 2.1}, we introduce 
a new object---$scalar  ~monopole$ in two-dimensional flat space
to represent the point defect there.
Using the scalar monopole,
we re-approach the usual Chern-Simons field theory of anyons 
in Subsection {\bf 2.2}, 
and show that the statistics of anyons is determined by the 
$charge$ of the scalar monopole.\\

\begin{center}
\section*{2.1. Scalar monopole in two dimensional space}
\end{center}

In two-dimensional flat space
${\bf x}=(x_{1}, x_{2})$, there is no magnetic field in the 
usual sense, since when charged particles move in a plane,
the induced magnetic field is always perpendicular to the plane.
Nevertheless, to describe the two-dimensional topology 
conveniently, we introduce $topological~
magnetic~ field$ from the purely mathematical consideration
as follows.

For simplicity, we  consider only the static case, 
therefore neglect the $time$ dimension.
Different from the  vector potential of magnetic
field in three-dimensional space, the potential 
of the topological magnetic field here
 is assumed to be  scalar, denoted as $a({\bf x})$.  
The topological magnetic field ${\bf b}=( b_{1}, b_{2})$ is
then defined as
\begin{equation}\left\{
\begin{array}{l}
 b_{1}=\frac{\partial}{\partial x_{2}}a, \\[.12in]
b_{2}=-\frac{\partial}{\partial x_{1}}a.
\end{array}\right.
\end{equation}
For a regular potential $a({\bf x})$  without singularity,  
the above  ${\bf b}({\bf x})$ satisfies the relation:  
\begin{equation}
{\bf \nabla}\cdot {\bf b}=0,
\end{equation}
which is consistant with the usual definition of magnetic field.
If we regard $a({\bf x})$ as a gauge field, we may study 
the gauge invariance of $a({\bf x})$: 
Let $a'({\bf x})=a({\bf x})+f$, if 
$ {\bf b}({\bf x})$ is invariant under this transformation, 
 following Eq. (1) we have
\begin{equation}
\frac{\partial f}{\partial x_{1}}= 
\frac{\partial f}{\partial x_{2}}=0,
\end{equation}
thus, $f$ can only be a constant 
(this holds for the static case only). 
 
It is evident that the topology in the above 
Eq. (2) is trivial.
To describe the nontrivial topology, we consider such a case 
that ${\bf b}({\bf x})$ is generated by a point-like source:
\begin{equation}
{\bf \nabla}\cdot {\bf b}= \delta({\bf x}).
\end{equation}  
Analogous to the definition of Dirac's magnetic
monopole in three-dimensional space, we see that Eq. (4)
also defines a topological monopole in a plane with
the monopole $charge$  equal to one.  One infers that
the potential $a({\bf x})$ in this case must be singular.
To solve the monopole, we turn to the method of Green function: 
Let $ {\bf b}={\bf \nabla} G$, equation (4) becomes
\begin{equation}
{\bf \nabla}^{2} G({\bf x})= \delta({\bf x}),
\end{equation}
which is solved as
\begin{equation}
G({\bf x})=\frac{1}{2\pi}\ln x,
\end{equation}  
where $x=|{\bf x}|$. Then 
\begin{equation}
{\bf b}({\bf x})=\frac{1 }{2\pi x^{2}}{\bf x}.
\end{equation}  
Using the relation between $a({\bf x})$ and $G({\bf x})$:
\begin{equation}
\partial_{i}a({\bf x})=-\epsilon_{ij}\partial_{j}G({\bf x}),
\end{equation}  
we obtain the scalar potential of the monopole Eq. (4) as
\begin{equation}
a({\bf x})=\frac{1 }{2\pi}\tan^{-1}
\left(\frac{x_{2}}{x_{1}}\right),
\end{equation}  
where the ``gauge" condition $f=0$ is chosen.
The above expression shows that $a({\bf x})$
is singular on the $x_{2}$-axis. For convenience, we
write $a({\bf x})$ into the polar coordinates $(r, \theta)$ as
\begin{equation}
a({\bf x})=\frac{1 }{2\pi}\theta({\bf x}).
\end{equation} 
In the meantime we should define a branch cut for 
instance $\theta=0$ to 
eliminate the ambiguity in the expression (10). 
Since the topological monopole in two-dimensional space is 
described by the scalar potential, 
we call it $scalar~ monopole$. 

In order to describe the above nontrivial topology
without concerning the singularity in potential,
we adopt an often-used quantity---$Winding~ number$,
which is defined through a loop $l$ as:
\begin{equation}
W=\oint_{l}~( \epsilon_{ij} b_{i} ) ~ dl_{j}.
\end{equation} 
For the loop without encircling a monopole,
we  use Eq. (1) to write the winding number as: 
$W=\oint_{l}\partial_{i}a({\bf x}) dl_{i}$,
which gives: $W=0$; 
For the loop encircling a scalar monopole as shown in Fig. 1 (a), 
we obtain $W=1$, and further $W=n$ 
for the loop in Fig. 1 (b). These results is easily understood
through a homotopic analysis:
A plane having a defect is
homotopic to a ring $S^{1}$. It is the homotopy group,
\begin{equation}
\Pi_{1}(S^{1})=Z,
\end{equation} 
that accounts for the winding number. 

As a topological invariant, $W$ is evidently 
independent of the definition of branch cut.
We now briefly show how to evaluate $W$ by taking 
into account the branch cut. Before doing this, 
we first introduce the following function,
which is related to the branch cut $\theta=0$,
\begin{equation}
\Delta ({\bf x})=\theta(-{\bf x})-\theta({\bf x})=
\left\{
\begin{array}{l}
\pi~ sign (x_{2}),  ~~~x_{2}\neq 0;\\[.12in]
\pi~ sign (x_{1}),~~~x_{2}= 0.
\end{array}\right.
\end{equation} 
We  divide the loop into two parts: $l_{1}$ and $l_{2}$ as 
shown in Fig. 1 (a). Since $l_{1}$ does not pass through the  
branch cut, we may use Eq. (13) to obtain:
\begin{equation}
W_{1}
=\int_{l_{1}}\partial_{i}a({\bf x}) dl_{i}=\frac{1}{2\pi}
\left[\theta(-{\bf x}')-\theta({\bf x}')\right]=\frac{1}{2}.
\end{equation} 
While $l_{2}$ pass through the cut branch anti-clockwise, 
we must add $2\pi$ to the integral,
\begin{equation}
W_{2}=\int_{l_{2}}\partial_{i}a({\bf x}) dl_{i}=\frac{1}{2\pi}
\left[2\pi+\theta({\bf x}')-\theta(-{\bf x}')\right]=\frac{1}{2}.
\end{equation}  
Sum up $W_{1}$ and $W_{2}$, we still get to $W=1$.

Since the scalar monopole is a mathematical
description of the nontrivial two-dimensional topology, 
we do not need to quantize it here.
We will show below that the scalar
monopole plays a crucial role in the statistics 
of identical particles in (2+1)-dimensional spacetime.\\

\begin{center}
\section*{2.2. Quantum field theory of anyons in 
(2+1)-dimensional spacetime}
\end{center}

Consider a system of nonrelativistic spinless fermion field
$\psi({\bf x},t)$ of mass $m$ and charge $e$, 
which minimally couples to
an Abelian Chern-Simons gauge field $A_{i}({\bf x},t)$.
The action is written as ($\hbar=c=1$)
\begin{equation}
S=\int
d^{3}x\left\{i\psi^{\dagger}D_{0}\psi+
\frac{1}{2m}\psi^{\dagger}{\bf D}^{2}\psi+
\frac{\lambda}{2}\epsilon^{\alpha\beta\gamma}
A_{\alpha}\partial_{\beta}A_{\gamma}\right\},
\end{equation}
where  $D_{\alpha}=\partial_{\alpha}+ieA_{\alpha},
$ ~ $F_{\alpha\beta}=\partial_{\alpha}A_{\beta}-
\partial_{\beta}A_{\alpha}$,~ $\lambda$
is a constant, and the metric is chosen to be $\eta=(+--)$.
The action is easily quantized. 
For example, for the fermion field, it is
\begin{equation}
\left\{\psi({\bf x},t),~ \psi^{\dagger}({\bf x}',t)\right\}
=\delta({\bf x}-{\bf x}').
\end{equation} 

Varying $S$ with respect to $A_{\alpha}$ gives:
\begin{equation}
\epsilon^{\alpha\beta\gamma}\partial_{\beta}A_{\gamma}(x)
=\frac{e}{\lambda}j^{\alpha}(x),
\end{equation} 
where the current
$j^{\alpha}$ are given by: $j^{0}=\rho=\psi^{\dagger}\psi$, 
$j^{i}=\frac{i}{2m}[\psi^{\dagger}D^{i}\psi-
(D^{i}\psi^{\dagger})\psi]$, $i=1,2,$ and they
satisfy the continuity equation:
$\partial_{\alpha}j^{\alpha}=0$.
Equation (18) implies that the Chern-Simons field is
determined by the particle current. This can be seen clearly
by the following calculations. 
 
We first look at the $\alpha=0$ component of equation (18):
\begin{equation}
\partial_{1}A_{2}(x)-\partial_{2}A_{1}(x)=
\frac{e}{\lambda}\rho(x),
\end{equation}
which is in fact a constraint equation. 
To solve $A^{i}$, $i=1,2$, we impose 
the Coulomb's gauge condition: $\partial_{i}A^{i}=0$,
which enables us to let $A^{i}=\epsilon^{ij}\partial_{j}\Omega$. 
Equation (19) then becomes:
\begin{equation}
{\bf \nabla}^{2}\Omega(x)=\frac{e}{\lambda}\rho(x).
\end{equation}
Employing the Green function as in Eq. (5), 
we can formally solve Eq. (20) and obtain further:
\begin{equation}
A^{i}(x)=\epsilon^{ij}\frac{\partial}{\partial x^{j}}
\left[\frac{e}{\lambda}\int d^{2}{\bf y} ~G({\bf x}
-{\bf y})\rho(y)\right].
\end{equation}
Using Eq. (8), the relation between Green function and 
the scalar monopole, we express the above $A^{i}(x)$ as the 
following compact form,
\begin{equation}
A^{i}(x)=\frac{\partial}{\partial x_{i}}
\Theta(x), 
\end{equation}
where
\begin{equation}
\Theta(x)=
\frac{e}{2\pi\lambda}\int d^{2}{\bf y} ~\theta({\bf x}
-{\bf y})\rho(y).
\end{equation}
One observes that the scalar monopole enters the
solution of  $A^{i}(x)$ naturally.
This makes it clear that the Chern-Simons gauge theory
embeds the nontrivial topological structure.
Notice that it must be sure that $\Theta(x)$ is 
a well-defined function when moving the derivative
out of $\Theta(x)$. There is no problem here,
since the density of fermion field $\rho(y)$
is a point-like function in the nonrelativistic case.

To solve $A_{0}(x)$, we look at the 
$\alpha=i,~i=1,2$ components of equation (18):
\begin{equation}
\partial_{i}A_{0}(x)-\partial_{0}A_{i}(x)=
-\frac{e}{\lambda}\epsilon_{ij}j^{j}(x).
\end{equation}
After a further derivative $\partial/\partial x_{i}$
acting on Eq (24) and using Coulomb's gauge condition, 
we turn Eq. (24) into a Laplacian equation. 
Then adopting the Green function, we solve $A^{0}(x)$ as
\begin{equation}
A_{0}(x)=-\frac{e}{\lambda}\epsilon_{ij}\int d^{2}{\bf y}~
G({\bf x}-{\bf y})\frac{\partial}{\partial y^{i}}j^{j}(y)
\end{equation}
Using the integration in parts and  the continuity 
equation of the current, we get to [See Appendix A]:
\begin{equation}
A_{0}(x)=\partial_{0}\Theta(x).
\end{equation}

Combining Eqs. (22) and (26), we arrive a remarkable 
conclusion that the Chern-Simons gauge field can be formally 
solved as a pure gauge: $A_{\alpha}(x)=\partial_{\alpha}
\Theta(x)$, where $\Theta(x)$ is the ${\bf x}$-site scalar 
potential of all the topological monopoles 
in two-dimensional space.

The above solution of the gauge field allows us to remove
the Chern-Simons gauge field from the action $S$
by the following  singular gauge transformations,
\begin{equation}\left\{
\begin{array}{l}
A'_{\alpha}(x)=A_{\alpha}(x)-\partial_{\alpha}
\Theta(x)=0,\\[.12in]
\psi'(x)=e^{ie \Theta(x)}\psi(x),\\[.12in]
\psi'^{\dagger}(x)=\psi^{\dagger}(x)e^{-ie \Theta(x)}.
\end{array}\right.
\end{equation} 
Under these transformations, the action $S$ turns out to be
\begin{equation}
S'=\int
d^{3}x~\left[i\psi'^{\dagger}\partial_{0}\psi'+
\frac{1}{2m}\psi'^{\dagger}{\bf \nabla}^{2}\psi'\right].
\end{equation}
Before discussing the physical meaning of the field $\psi'(x)$
from the above action, we first look at the function $\Theta(x)$.  
From the expression (23), we know that $\Theta(x)$ 
represents the total potential of the scalar
monopoles with charge density $\frac{e}{2\pi\lambda}\rho(y)$ 
in the whole space as shown in Fig. 2 (a). 
Moreover,  the equation
\begin{equation}
\left[\Theta({\bf x}, t), ~\psi'({\bf x}', t)\right]
=-\theta({\bf x}-{\bf x}')\psi'({\bf x}',t),
\end{equation} 
indicates that even if $\Theta({\bf x}, t)$ and 
$\psi'({\bf x}', t)$ are space-likely separated,
they are still related with each other. Therefore,
 $\Theta(x)$ is essentially a nonlocal quantity.
We further infer that the new field 
$\psi'(x)$ is a nonlocal field that combines the
topological monopoles in whole space with the fermi
field at {\bf x}. In this sense, $\psi'(x)$ describes a set 
of quasi-particles only, which is far from free, 
though these quasi-particles have the free-form action (28).
In the presence of topological monopoles, 
the statistics of these quasi-particles differs from the 
that of the fermions. A simple calculation by 
using Eqs. (17) and (27) gives
\begin{equation}
\psi'(x)~\psi'(x')
=-e^{-i\kappa \Delta({\bf x}-{\bf x}')}
~\psi'(x')~\psi'(x), ~~~~t=t',
\end{equation} 
where  $\kappa=e^{2}/\lambda$. The multi-valued function
$\Delta({\bf x}-{\bf x}')$ is defined in Eq. (13), 
its values ($\pm\pi)$ are specified
by the relative positions of ${\bf x}$ and ${\bf x}'$.
The above commutation rule is understood by 
the way drawn in Fig. 2.b that 
when two quasi-particles exchange their positions, 
one of them must  pass through the monopole of each others, 
this will result in 
a nontrivial phase factor in terms of the monopole charge
($e/\lambda$) that determines statistics of the quasi-particles.
In general,  the nonlocal field $\psi'(x)$ obey an anyonic 
statistics interpolated between bosons and fermions.
It should be emphasized that even if $\kappa$
is an add number, namely, $\psi'(x)$ satisfies a bosonic 
commutation relation, $\psi'(x)$ can not be taken as a boson 
field, it still describes the quasi-particles, 
and does not commute with the fermion field  $\psi(x)$. 
Likewise,  $\psi'(x)$ is not the real fermion field for 
the even  $\kappa$.

We conclude that the anyon field  embeds
not only the property of nonlocality in
physical interest,
but also the nontrivial two-dimensional topology.
It is the topological monopole that gives rise
to the exotic statistics of the anyon field.\\

\begin{center}
\section*{2.3.  Chern-Simons field as a realization of 
Berry's potential }
\end{center}

Consider a quantum mechanical system in which the 
Hamiltonian $H$ evolves adiabatically with parameters 
$R_{i}\equiv R_{i}(t)$, and  has discrete eigenvalues 
$ E_{n}({\bf R})$:
\begin{equation}
H({\bf R})|\Psi_{n}({\bf R}) \rangle = E_{n}({\bf R})
|\Psi_{n}(R{\bf R}) \rangle,
\end{equation} 
where $|\Psi_{n}({\bf R}) \rangle$ are eigenstates.
If ${\bf R}$ executes a closed loop in  parameter space: 
${\bf R}(0)={\bf R}(T)$, Berry  \cite{berry} proved that after 
a round trip along the loop, the state of the system will 
acquire a geometric phase: $\gamma=\oint A_{i}({\bf R}) d R^{i}$, 
which is attributed to the holonomy in the parameter space,
where 
\begin{equation}
A_{i}({\bf R})\equiv i\langle\Psi_{n}({\bf R})|\frac{\partial}
{\partial R^{i}} |\Psi_{n}({\bf R}) \rangle,
\end{equation} 
is called Berry's potential and has wide 
application in physics.

It is well known that the vector potential of Dirac's 
magnetic monopole in three-dimensional space can be realized by
Berry's potential of the system of spin-$1/2$ particle 
moving in the magnetic field which traces out a circuit 
slowly without changing the magnitude \cite{berry}.
In two-dimensional space, the object analogous to Dirac's
magnetic monopole is evidently the Chern-Simons gauge field. 
Choosing the classical limit to the density of fermions in  
Eq. (21): $\rho(x)=\delta({\bf x})$, we get to the usual 
expression of the Chern-Simons field for single particle:
\begin{equation}
A_{i}({\bf x})=\frac{e}{2\pi\lambda}
\frac{\epsilon_{ij}x^{j}}{{\bf x}^{2}}.
\end{equation} 
One may ask: Can the Chern-Simons 
field be realized as Berry's potential?

Since there is no the usual magnetic field in two spatial 
dimensions, we have to turn to the topological magnetic 
field introduced in Subsection {\bf 2.1}. 
Then the above question becomes the purely mathematical scheme.

Let us consider a constant topological magnetic field
${\bf b}=(b_{1},~b_{2})$, the corresponding scalar 
potential is simply: $a({\bf x})=b_{1}x_{2}-b_{2} x_{1}$. 
We notice that this magnetic field can also be 
realized as the long distance approximation of the 
topological monopole Eq. (7). Assuming 
a spin-$1/2$ particle moving in this field, 
the simplified Hamiltonian is
\begin{equation}
H=\eta(\sigma_{1} b_{1}+\sigma_{2} b_{2}),
\end{equation} 
where $\sigma_{1,2}$ are two Pauli matrices, $\eta$ 
accounts for the magnetic  moment and Lander factor.
The eigenkets of the Hamiltonian for
spin up and down are respectively,
\begin{equation}
|\phi_{1}\rangle=\frac{1}{\sqrt{2}}\left(
\begin{array}{c}
1\\
e^{i\theta}
\end{array}\right), ~~~
|\phi_{2}\rangle=\frac{1}{\sqrt{2}}\left(
\begin{array}{c}
1\\
e^{-i\theta}
\end{array}\right),
\end{equation} 
where $\theta=\tan^{-1}(b_{2}/b_{1})$.
For the case that the magnetic field ${\bf b}=(b_{1},~b_{2})$, 
taken as the parameter space, traces out a circuit 
slowly without changing the magnitude, we use Eq. (32) to 
obtain the Berry's potential
for the ket for instance $|\phi_{1}\rangle$ as
\begin{equation}
A_{i}'({\bf b})= i\langle\phi_{1}|\frac{\partial}{\partial b_{i}}
|\phi_{1}\rangle=\frac{1}{2}\frac{\epsilon_{ij}b_{j}}{{\bf b}^{2}}.
\end{equation} 
The above result is identified with 
$A_{i}$ in Eq. (33) up to a constant factor.
 \\

\begin{center}
\section*{III.~ Topology and statistics in 
(3+1)-dimensional spacetime}
\end{center}

In last section we have shown that topology determines
the statistics of identical particles in two-dimensional 
flat space. The naive model Eq. (34) indicates that 
the Chern-Simons field is analogous 
to the vector field of Dirac's monopole from the viewpoints
of both topology and Berry's potential. 
This observation inspires us to investigate further the role  
topology plays in three-dimensional statistics.

In this section, we first recapitulate the conventional
conclusion of that there are only bose and fermi 
statistics in (3+1)-dimensional spacetime, and point out that
this conclusion essentially results from the global and 
relatively trivial topology. We then present a simple 
model of anyons in three-dimensional space,
in which infinite number of Dirac's magnetic monopoles 
appear, they change the topology, and give rise to 
the exotic statistics \cite{monopole}.\\

\begin{center}
\section*{3.1. Statistics in  trivial topology }
\end{center}

It was believed that identical particles in 
three-dimensional space follow either bose or fermi statistics.
This belief  generally comes from the following two arguments: 

(i)  The non-Abelian nature of three-dimensional Rotation group 
implies that its representation is labeled by a discrete 
index, integer or half-integer angular momentum. It was 
proved \cite{pauli} by the local 
relativistic quantum field theory 
that the particles having integer angular momenta 
obey the bose statistics, and the particles having
half-integer angular momenta obey the fermi statistics. 
However, the proof becomes invalid
in the case of nonrelativistic spinless fields, especially for
the quasi-particles described by $nonlocal$ fields.

(ii) Another argument is based on topology.
In three-dimensional Euclidean space 
$\Re^{3}$, a closed curve encircling a singularity twice 
can be continuously contracted to a point without passing 
through the $singularity$  \cite{leinaas}.
Therefore, for a system of two identical
 particles at ${\bf x}_{1}$, ${\bf
x}_{2}$, the  wave function
$\Psi({\bf x}_{1}, {\bf x}_{2})$ should satisfy 
\begin{equation}
\hat{P}_{x}^{2}\Psi({\bf x}_{1}, {\bf x}_{2})= 
\Psi({\bf x}_{1}, {\bf x}_{2}),
\end{equation}
where $\hat{P}_{x}$ is the permutation operator  
defined by the parallel displacement, 
and the point  
\begin{equation}
{\bf x}={\bf x}_{1}-{\bf x}_{2}=0
\end{equation}
is taken as the $singularity$. Following Eq. (37), 
one gets to that $\hat{P}_{x}$ has constant eigenvalues  
$p=\pm 1$ only, corresponding to bose and fermi cases, 
respectively.

In fact, the above argument (ii) 
can be alternatively achieved through a homotopic 
analysis of $\Re^{3}$. For this purpose, we still
adopt the above example of two identical particles at
${\bf x}_{1}$ and ${\bf x}_{2}$.  Excluding the
singularity as in Eq. (38) in the state of 
the system, $\Re^{3}$ turns out to be \cite{leinaas}:
\begin{equation}
\Re^{3}-\{0\}\approx(o,
\infty)\times \wp^{2},
\end{equation}
where $ \wp^{2}$ is the two-dimensional projective plane 
for the direction $\pm{\bf x}/|{\bf x}|$ of ${\bf x}$, 
it is also topologically equivalent to a doubly connected
surface of a three-dimensional sphere with diametrically 
opposite points identified. 
Then the homotopy group of the system becomes:
\begin{equation}
\Pi_{1}(\wp^{2})=\Pi_{1}\left[SO(3)\right]=Z_{2}.
\end{equation}
This result was frequently used to interpret 
that there are only bose and fermi statistics in the 
ordinary three-dimensional space \cite{finkelstein,zee}.

It is evident that the above argument (ii)
(also as those in Refs. [3] and [19])
is based on the global topology, which does not concern 
the concrete structure of the system.
The singularity shown in Eq. (38) is independent of the system 
itself, it comes from the permutation of two particles only. 
However, for the two-dimensional anyon system in last section, 
the nontrivial topology (the scalar monopole) is not 
caused by permutation of two anyons, but by the Chern-Simons 
term, which determines the permutation rule as the result. 
Further, there are infinite number of scalar monopoles 
associate with the anyon field in its construction as Eq. (27).

The above discussion makes it clear that the 
usual argument of three-dimensional statistics is based on
the relatively trivial topology, compared with the scalar 
monopole of two dimensions. In other words, the usual 
conclusion holds only in the case that there is no other
nontrivial topological singularity involved in the system. \\

\begin{center}
\section*{3.2. A simple model for exotic statistics in 
(3+1)-dimensional spacetime}
\end{center}

In this section we attempt to investigate how the topological
singularity for example magnetic monopole affects the 
statistics of a physical system \cite{monopole}. For the 
purpose, we construct a model by a system of non-relativistic 
spinless fermionic matter field $\psi(x)$ of mass $m$ and 
charge $e$, which couples to an Abelian gauge field $A_{\mu}(x)$.
The gauge field $A_{\mu}(x)$interacts with a scalar 
field $\chi(x)$ through the Pontrjagin term.  Without loss
of generality, the action is: 
\begin{equation}
S=\int
d^{4}x[i\psi^{\dagger}D_{0}\psi+\frac{1}{2m}
\psi^{\dagger}{\bf D}^{2}\psi+
\frac{e^{2}}{16\pi^{2}}\epsilon^{\mu\nu\sigma\rho}
F_{\mu\nu}F_{\sigma\rho} ~\chi].
\end{equation}
Since the dynamical term of $\chi(x)$ is not included in action,
$\chi(x)$ is taken as an external source, 
its physical implication will be discussed later. 

Before studying the above action, 
we first present two remarks on it:

(i)~ The term
$\L_{0}(x)=\frac{e^{2}}{16\pi^{2}}\epsilon^{\mu\nu\sigma\rho}
F_{\mu\nu}F_{\sigma\rho}
\chi$ in the action is not a purely topological
invariant since $\chi(x)$ is a spacetime-dependent function. 
$\L_{0}(x)$ has been extensively used in
quantum field theory to deal with the problems
such as the  CP violation in strong 
interaction \cite{witten} and $\theta$-vacuum \cite{poly}, etc.
However, the physical meaning of 
the gauge field $A_{\mu}(x)$ in our model is
essentially  different from that in Refs. [21, 22]:
$A_{\mu}(x)$ here is not the true electromagnetic potential,
it is entirely determined by the particle current and
$\chi(x)$ (see the following Eq. (42)), therefore has no 
independent dynamics. In this sense, $A_{\mu}(x)$ 
is similar to the Chern-Simons field of the anyons, whose
purpose in life is to implement exotic statistics to particles
\cite{chen}.

(ii) The above action is different from the one
used by Goldhaber $et ~al.$ in Ref. [23], where they 
showed that the Pontrjagin term ( $\chi$ is a
constant in $\L_{0}(x)$) does not contribute to
the statistics of particles, since it
is a total derivative \cite{instanton}, and vanishes
after integrating over the gauge field. 
Moreover, the gauge field in  Ref. [23] has Maxwell-like 
dynamics, which does not appear in our model.

To deal with the above action, we  follow the procedure
of treating the action (16) in  Section {\bf II}. 
The Lagrangian equation of $A_{\mu}$ is obtained to be: 
\begin{equation}
\epsilon^{\mu\nu\sigma\rho}
F_{\sigma\rho}\partial_{\nu}\chi=4\pi^{2} e^{-1}j^{\mu},
\end{equation}
where the current $j^{\mu}$ are given by: $j^{0}=\psi^{\dagger}
\psi=\rho$, $j^{i}=\frac{i}{2m}[\psi^{\dagger}D^{i}\psi-
(D^{i}\psi^{\dagger})\psi]$, and they satisfy the continuity
equation: $\partial_{\mu}j^{\mu}=0$. For convenience, we further 
separate Eq. (42) into 
two equations with respect to $\mu=0$ and $\mu=i= 1,2,3$:
\begin{equation}
\partial_{i} C^{i}(x)=4\pi^{2} e^{-1}\rho(x),
\end{equation}
\begin{equation}
-\partial_{0}C^{i}(x)-\epsilon^{ijk}\partial_{j}
D_{k}(x)=4\pi^{2}e^{-1} j^{i}(x),
\end{equation}
where 
\begin{equation}
C^{i}(x)=\epsilon^{ijk}\partial_{j} A_{k}(x)\chi(x),
 ~~~~D_{i}(x)=\left[\partial_{i} A_{0}(x)-
\partial_{0} A_{i}(x)\right]\chi(x).
\end{equation}
Equation (43) is in fact a constraint among the charge 
density, gauge flux and the scalar field. In the lattice 
formulation, equation (43) represents a duality transformation
between the current in the lattice and field strength in 
the dual lattice. Our calculation is carried in the continuous
space, it is easily copied to the lattice. 

To solve the vector component of the gauge field,  
we choose the subsidiary condition:
$\partial_{i} A^{i}=0$,  and let $ A_{i}=
\epsilon_{ijk}\partial^{j}\Phi^{k}$, 
while  $\partial_{i}\Phi^{i}=0$.  
We then obtain from Eq. (45) that: 
\begin{equation}
\nabla^{2} \Phi^{i}(x)
=-\frac{ C^{i}(x)}{\chi(x)},
\end{equation}
a Laplacian equation,  which can be solved by 
using the Green function
$G({\bf x}-{\bf x}')$:  $\nabla^{2}G({\bf x}-{\bf x}')=
\delta({\bf x}-{\bf x}')$, and gives further:  
\begin{equation}
A_{i}(x)=-\int d^{3}{\bf x}^{\prime}~
\epsilon_{ijk}\frac{\partial}{\partial x_{j}}G({\bf
x}-{\bf x}')\left[\frac{ C^{k}(x')}{\chi(x')}\right]~.
\end{equation}

To simplify Eq. (47), we introduce  the
vector potential  of Dirac's 
magnetic monopole ${\bf M}$ by the equation
\begin{equation}
\frac{\partial}{\partial x_{j}}
G({\bf x}-{\bf x}')=\epsilon^{jkl}\frac{\partial}{\partial x^{k}}
M_{l}({\bf x}-{\bf x}').
\end{equation}
Notice that there are many 
${\bf M}$ satisfying Eq. (48), and they are
connected with each other by gauge transformations.  
With the help of Eq. (48), we turn Eq. (47) into
\begin{equation}
 A_{i}(x)=\int d^{3}{\bf x}^{\prime}
\left[\frac{\partial}{\partial x^{i}} M_{j}({\bf x}-
{\bf x}')~\frac{C^{j}(x')}{\chi(x')}-
\frac{\partial}{\partial x^{l}} M_{i}({\bf x}-{\bf
x}')~\frac{C^{l}(x')}{\chi(x')}\right].
\end{equation}
We observe that $\frac{\partial}{\partial x^{l}}
M_{i}({\bf x}-{\bf x}')=-\frac{\partial}
{\partial x'^{l}} M_{i}({\bf x}-{\bf x}')$.
After a integration in parts and using Eq. (45), 
we obtain that second term in the right of 
Eq. (49) vanishes. Then,
\begin{equation}
 A_{i}(x)= \partial_{i}\Omega(x),
\end{equation}
where
\begin{equation}
\Omega(x)\equiv
\int d^{3}{\bf x}^{\prime}~M_{j}({\bf x}-{\bf
x}')~\frac{C^{j}(x')}{\chi(x')}.
\end{equation}
Using  Eq. (45),
one can prove that $\Omega(x)$ is invariant under the 
transformation 
\begin{equation}
M'_{i}({\bf x})= M_{i}({\bf x})-
\partial_{i}\beta({\bf x}), 
\end{equation}
which implies that the above results are principally
independent of a particular  ${\bf M}$.

We now solve $A_{0}(x)$ by Eq. (45). 
A left-derivative acting
on $D(x)/\chi(x)$ gives:
\begin{equation}
\nabla^{2}A_{0}(x)=\frac{\partial}{\partial x_{i}}
\left[\frac{D_{i}(x)}{\chi(x)}\right].
\end{equation}
Employing the Green function, we obtain $A_{0}(x)$ as
\begin{equation}
A_{0}({\bf x})=\int d^{3}{\bf x}^{\prime} G({\bf x}-
{\bf x}')~ \frac{\partial}{\partial x'_{i}}
\left[\frac{D_{i}(x')}{\chi(x')}\right]
\end{equation}
Using Eqs. (44) and (45), and integration in parts, 
we get to [See Appendix B]:
\begin{equation}
A_{0}({\bf x})=\partial_{0} \Omega(x).
\end{equation}
Notice that some techniques adopted in the above calculations 
such as integral in part, removing the derivative
$\partial_{i}$ out of Eq. (49)  to obtain Eq. (50),   
are feasible by checking the behaviors of 
quantities ${\bf M},~{\bf C}/\chi$, etc., around the origin
${\bf x}=0$ and in long distance as well. 
Since ${\bf M}$ is singular, one has to assume a branch cut 
in space, which will be discussed later.

Combining Eqs. (50) and (55) into a covariant expression: 
$A_{\mu}(x)=\partial_{\mu}\Omega(x)$,
we arrive an conclusion that $A_{\mu}(x)$  is 
formally solved as a pure gauge, 
$albeit$ of a nonstandard form, because of 
the singularity (or multi-valued) in 
the vector potential of magnetic  monopole. 

It is interesting to notice that the way of the Dirac's 
magnetic monopole's entering the gauge field $A_{\mu}(x)$
is exactly the same  way as the scalar monopole's entering 
the Chern-Simons gauge field in two-dimensional space.
This similarity leads to the following analysis
about the three-dimensional statistics.

Let us return to the action
$S$ as in Eq. (41). Since 
$A_{\mu}$ is formally solved as a pure gauge, 
it can be removed from the action through a
peculiar gauge transformation,
\begin{equation}
A'_{\mu}(x)=A_{\mu}(x)-\partial_{\mu}\Omega(x)=0,
\end{equation}
The corresponding  transformations to the fermion field are 
\begin{equation}
\varphi(x)=e^{ie\Omega(x)} \psi(x), ~~~\varphi^{\dagger}(x)=
\psi^{\dagger}(x)e^{-ie\Omega(x)}.
\end{equation}
By Eqs. (56) and (57),  $S$ is transformed into the 
following free-form:
\begin{equation} 
S=\int
d^{4}x[i\varphi^{\dagger}\partial_{0}\varphi+
\frac{1}{2m}\varphi^{\dagger}\partial_{i}^{2}\varphi].
\end{equation}
However, $\varphi(x)$ is far from a free field.
In order to understand the physical
meaning of  $\varphi(x)$, we first make some analysis to
the function $\Omega(x)$:

From Eq. (51), we know that $\Omega(x)$ represents the vector 
potential of Dirac's magnetic monopoles 
in the whole space along the direction ${\bf C}(x')$,
where the charge density of the monopoles is ${\bf C}(x')
/\chi(x')$. Thus,  similar to the function
$\Theta (x)$ in Eq. (23), $\Omega(x)$ is a nonlocal quantity.
Further,  equation (50) indicates that the term $\partial_{i}
\Omega(x)$ is not a trivial quantity: The vector field
${\bf A}(x)$ describes a point-like ``magnetic" flux localized at 
${\bf C}(x)/\chi( x)$, where ${\bf C}(x)$ is associated with  
particle density, and the ``magnetic" flux is measured through 
the Aharonov-Bohm effect:
\begin{equation}
\int_{{\bf \partial}\omega}d\vec{l}\cdot{\bf A}=
\int_{\omega}d\vec{\omega}\cdot\frac{{\bf C}}{\chi},
\end{equation}
where $ \partial \omega$ is the boundary of area $\omega$.
These facts suggest that $\Omega(x)$ seems to be
the three-dimensional duplication of $\Theta (x)$.
However, they are quite different:  In two-dimensional
space, when the charged particles move 
in a plane, the induced ``magnetic" flux is always
perpendicular to the plane, thus it can be treated as a scalar
and directly associated with the particle density as shown 
in Eq. (19); In three-dimensional space, we have to adopt 
a ``bridge" to connect the vector--``magnetic" 
fluxes to the scalar--particle density. 
The scalar field $\chi(x)$ in $S$ is really a simple 
(also cheap) ``bridge" that works in Abelian gauge theory.

From the above discussions, we infer that $\varphi(x)$,
constructed by $\Omega(x)$, is essentially a nonlocal 
field, which describes a system of quasi-particles
composed of fermions and infinite number of magnetic
monopoles in space. 
Therefore, when two quasi-particles exchange
their positions, one of them must pass through the monopole
of others. This permutation will 
produce an unusual phase factor in terms of magnetic monopoles,
which leads to  that the field $\varphi(x)$ follows a 
different commutation rule from that of the fermion 
field $\psi(x)$. Choosing a particular source function 
$\chi(x)$, one can compute the commutation rule of 
$\varphi(x)$ using the following equation which 
is proved in Appendix C,
\begin{equation}
\frac{\partial\chi(x)}{\partial x_{i}}
\left[\frac{ C_{i}(x)}{\chi( x)}, 
~~\psi(x')\right]_{t=t'}=-\frac{4\pi^{2}}{e}\psi(x')
~\delta({\bf x}-{\bf x}').
\end{equation}

We now present an explicit example to illustrate that 
$\varphi(x)$ generally obeys an exotic commutation rule.  
To avoid the singularity in the vector potential of 
magnetic monopole,
we choose the vector potential as in Ref. [24], 
\begin{equation}\left\{
\begin{array}{l}
{\bf M}_{1}=(4\pi r)^{-1}\tan(\frac{\theta}{2})~{\bf \hat{ 
\phi}},~~ ~~~~0\leq\theta<\frac{\pi}{2}+\delta_{1},\\[.13in] 
{\bf M}_{2}=-(4\pi
r)^{-1}\cot(\frac{\theta}{2})~{\bf \hat{  \phi}},~~~~
\frac{\pi}{2}-\delta_{2} <\theta\leq\pi.
\end{array}\right.
\end{equation}
In the overlap region $\frac{\pi}{2}-\delta_{2}
<\theta<\frac{\pi}{2}+\delta_{1}$, ${\bf M}_{1},$
and ${\bf M}_{2}$ are connected by the transformation:
${\bf M}_{2}={\bf M}_{1}+\nabla\beta$, where $\beta=-\phi/2\pi$.
Notice that this $U(1)$ transformation  does not require to
quantize the monopole charge, because the field
$\varphi(x)$ is invariant under this local transformation. 
Along with above choice of ${\bf M}({\bf x})$,
we need the ${\bf \hat{\phi}}$ component of ${\bf C}(x)$.
This can be achieved by choosing 
\begin{equation}\chi=g(\phi)\neq 0,
\end{equation}
where $\partial_{\phi}g(\phi)\neq 0$.  
Equation (60) is then reduced to be
\begin{equation}
\left[\frac{C_{\phi}(x)}{\chi( x)},
~~\psi(x')\right]_{t=t'}=-\frac{4\pi^{2}}{e}h({\bf x})\psi(x')
~\delta({\bf x}-{\bf x}'),
\end{equation}
where $h=r \sin\theta/\partial_{\phi}g(\phi)$.

Unfortunately, the above particular choice of $\chi( x)$
breaks the $rotation$ symmetry of the system,
which can be seen from Eq. (63). Moreover, there  
raises a problem in the meantime that the 
local $U(1)$ symmetry between ${\bf M}_{1}$ and ${\bf M}_{2}$
is broken when we compute the commutation role of 
the fields $\varphi(r)$ 
in the overlap region $\frac{\pi}{2}-\delta_{2}
<\theta<\frac{\pi}{2}-\delta_{2}$. 
This $U(1)$ symmetry breaking leads to a headache 
ambiguity to the commutation rule in the overlap region.
To resolve this ambiguity, we set up a $branch$ 
$cut$ for instance $x$-axis in three-dimensional space, then 
the whole space is divided into two regions 
$R_{1}$ and $R_{2}$ as shown in  Fig. 3:
\begin{equation}
R_{1}=\left\{
\begin{array}{ll}
\theta \in [0, \pi/2),& \phi\in [0, 2\pi], \\[.13in]
\theta=\frac{\pi}{2}, & \phi\in [0, \pi),
\end{array}
\right.
~~~~~R_{2}=\left\{
\begin{array}{ll}
\theta\in (\pi/2, \pi], &\phi\in [0, 2\pi], \\[.13in]
\theta=\frac{\pi}{2},& \phi\in [\pi, 2\pi).
\end{array}
\right.
\end{equation}
We require that ${\bf M}_{2}$ be restricted in 
$ R_{1}$ and ${\bf M}_{2}$ in $ R_{2}$, 
instead of the regions defined in Eq. (61).

With the above choices of  ${\bf C}$ and ${\bf M}$, we 
can compute the permutation rule of $\varphi(r)$ 
and obtain finally: (The symbol ${\bf x}$ is replaced
by ${\bf r}$ in follows)
\begin{equation}
\varphi({\bf r}_{1}, t) \varphi({\bf r}_{2},t)=
-\varphi({\bf r}_{2},t)\varphi({\bf r}_{1},t)
~e^{i\Gamma({\bf r}_{1}, ~{\bf r}_{2})},~~~~ {\bf r}_{1}\neq
{\bf r}_{2},
\end{equation}
where
\begin{equation}
\Gamma({\bf r}_{1},~{\bf r}_{2})=\left\{
\begin{array}{cl}
\pi r_{12}^{-1}\tan(\frac{\theta_{12}}{2})~
\left[h({\bf r}_{1})+h({\bf r}_{2})\right],
&~~{\bf r}_{12}\in R_{a},\\[.22in]
-\pi r_{12}^{-1}\cot(\frac{\theta_{12}}{2})~
\left[h({\bf r}_{1})+h({\bf r}_{2})\right],
&~~{\bf r}_{12}\in R_{b},
\end{array}
\right.
\end{equation}\\
where $r_{12}=|{\bf r}_{1}-{\bf r}_{2}|$, $\theta_{12}=
\arg({\bf r}_{12}, {\bf n}_{z})$. 
It follows from the relation 
$\theta_{12}=\pi-\theta_{21}$ that
\begin{equation}
\Gamma ({\bf r}_{1},~
{\bf r}_{2})=-\Gamma ({\bf r}_{2},~{\bf r}_{1}),
\end{equation}
which is consistent with the permutation rule (65).
Equation (65) explicitly shows that the new 
field $\varphi(x)$ obeys an exotic statistics, 
which is not confined in the domain of fractional 
statistics, but becomes 
a space-dependent functional statistics determined by  
$\Gamma({\bf r}_{1}, ~{\bf r}_{2})$. $\varphi(x)$ is 
still called  $anyon$ field.

We see that the three-dimensional rotation 
symmetry is inevitably sacrificed in the above example, 
which has similarly appeared in Ref. [17]. In this 
consideration, our model is only appropriate to the  
$condensed ~ matter$ system, where the $SO(3)$ 
symmetry is not necessarily preserved in many cases.
Certainly, one can choose other $\chi(x)$ 
to obtain the permutation rules  different from Eq. (65). 
This flexibility of $\chi(x)$ exceeds our usual 
understanding of the exotic statistics, and adds a new variable
to the system as well.
For the case that 
\begin{equation}
\chi=\epsilon (z)=\left\{
\begin{array}{l}
1, ~~~z\geq 0,\\
0,  ~~~z< 0,
\end{array}
\right.
\end{equation}
the system [mainly Eq. (42)] is reduced into
(2+1)-dimensional case discussed in Section {\bf II}, where the
Green function and topological monopole are different
from those in three-dimensional space.
This reduction has been pointed out by Wilczek in Ref. [25].

The wave functions of anyons
here can not be imitated by the  Laughlin's 
expression \cite{laugh}.
However, one can easily write down the state for a
many-anyon system in the Fock space. 

The equivalence of two actions [Eqs. (41) and (58)]
leads to a conclusion: Anyons in four dimensional 
spacetime can be realized through a system
of fermion field, Abelian gauge field and scalar field;  
It is infinite number of magnetic monopoles embodied in the 
anyon field that bring in system the nontrivial
topology, and give rise to the exotic statistics of 
anyons.  \\

\begin{center}
\section*{3.3. Physical meaning of the scalar field $\chi$}
\end{center}

In above Subsection {\bf 3.2} we have seen that the 
scalar field $\chi(x)$ plays an important role in the model 
of anyons. There it is interpreted as the external 
source function.  In this subsection, we hope to 
understand deeply the implication of $\chi(x)$ in our model
by approaching a chiral system. 

For simplicity in calculation, we consider a system of
relativistic massless fermion field $\psi'(x)$, which 
minimally couples to a gauge field $A_{\mu}(x)$.
The chirality is introduced by the 
interaction of  axial chiral current with the chiral mode
$\chi'(x)$ of the system. The effective action is
\begin{equation}
S_{eff}=\int
d^{4}x~\left\{\bar{\psi'}\left[i\gamma^{\mu}
(\partial_{\mu}+ieA_{\mu})\right]\psi'+
\partial_{\mu}\chi'j_{5}^{\mu}\right\},
\end{equation}
where $\gamma_{\mu}, \mu=0,1,2,3 $ are gamma matrices,
$j_{5}^{\mu}=\bar{\psi'}\gamma^{\mu}
\gamma_{5}\psi'$ are the axial chiral currents,
and the dynamical term of the chiral mode is neglected. 
It is easily proved that $S_{eff}$ is invariant 
under the local $U(1)$ chiral transformations:
\begin{equation}
\left\{
\begin{array}{l}
 \psi''=e^{i\alpha\gamma_{5}}\psi',\\[.12in]
\bar{\psi''}=\bar{\psi'}
e^{i\alpha\gamma_{5}},\\[.12in]
\chi''=\chi'-\alpha,
\end{array}
\right.
\end{equation}
where $\alpha\equiv\alpha(x)$ is an infinitesimal 
real parameter. In the case of low energy, 
we can safely integrate over the fermion fields  
in functional integral. $S_{eff}$ then becomes:  
\begin{equation}
S'_{eff}=\int 
d^{4}x~\ln\left\{ \det\left[i\gamma^{\mu}(\partial_{\mu}
+ie A_{\mu})+ \partial_{\mu}\chi'\gamma^{\mu}
\gamma_{5}\right]\right\}.
\end{equation}
By the perturbation theory, we expand $S_{eff}$ 
in powers of  $A_{\mu}$. 
To the second order, there are two non-vanishing terms:  
\begin{equation}
S_{1}^{(2)}=\frac{e^{2}}{4}\int d^{4}(xy)~
\langle j^{\mu}(x)j^{\nu}(y)\rangle~
A_{\mu}(x)A_{\nu}(y),
\end{equation}
\begin{equation}
S_{2}^{(2)}=-\frac{e^{2}}{4}\int
d^{4}(xyz)~\langle\partial_{\lambda}j_{5}^{\lambda}(x)
j^{\mu}(y)j^{\nu}(z)\rangle
~\chi'(x) A_{\mu}(y)A_{\nu}(z),
\end{equation}
where $j^{\mu}=\bar{\psi'_{s}}\gamma^{\mu}\psi'_{s}$. 
The first term $S_{1}^{(2)}$ is a Maxwell-like term, 
which can be canceled by adding the
same term but with opposite sign to the action $S_{eff}$. 
Our interest is in the second term
$S_{2}^{(2)}$. At one-loop level, there appear triangle graphs
that break down the usual Ward-Takahashi identity of the  
chiral current $ j_{5}^{\mu}$, and  leads to the triangle 
anomaly \cite{schwinger}. A detailed calculation gives
\begin{equation}
S_{2}^{(2)}=\frac{e^{2}}{16\pi^{2}}~
\epsilon^{\mu\nu\sigma\rho}\int
d^{4}x ~F_{\mu\nu} F_{\sigma\rho}\chi'.
\end{equation}
This result can be alternatively obtained using
the technique introduced by Fujikawa \cite{fuji}. 

Comparing the above term $S_{2}^{(2)}$ with the term $\L_{0}(x)$ 
in the action $S$ in Eq.(41), we identify immediately 
the scalar field $\chi(x)$ in $S$ with the chiral mode
$\chi'(x)$ in $S_{eff}$. These results imply that
the anyons in (3+1)-dimensional spacetime are originated from the 
chiral structure of the quantum system  as Eq.(69).
It is interesting to recall that the anyons 
in (2+1)-dimensional spacetime are associated with a chiral 
spin system \cite{wen,fradkin2}, where the chiral mode
is simply the mass of particles.
This similarity of anyons in different dimensions 
suggests that there presumably exists a deeper connection 
among chirality, topology and statistics, which is still 
unclear. \\

\begin{center}
\section*{IV.~ Conclusion}
\end{center}

In summary, we have discussed the connection between 
topology and statistics in two and three-dimensional
space, respectively.
With the introduction of scalar monopole in two-dimensional
space, we re-interpret the usual anyons, described by 
the Chern-Simons gauge field theory, 
as the quasi-particles constructed by 
fermions and the scalar monopoles in such a way that each fermion 
is surrounded by infinite number of scalar monopoles in 
whole space. By this quasi-particle 
picture, we observe that when two anyons exchange their positions,
each of them  must pass through the monopole of others. 
This exchange  will lead to a nontrivial phase factor in 
term of monopole $charge$. It is the monopole
charge that determines the statistics of the anyons.
 
By re-analyzing the conventional arguments
about the connection between topology and statistics 
in three spatial dimensions, we find that the usual conclusion 
of the bose and fermi statistics as the only possibilities 
there are based on the global topology, which is relatively trivial
compared with the scalar monopole of two spatial dimensions.
We construct a simple model of three-dimensional anyons 
by a system of non-relativistic fermion field, 
Abelian gauge field, and a scalar field 
which is taken as the external source. 
The Dirac's magnetic monopole enters the solution of 
the gauge field and changes the three-dimensional topology.
As a result, the anyon field is formulated by the fermion 
field and  Dirac's magnetic monopoles in such
a form that each fermion are surrounded by 
infinite number of magnetic monopoles in the whole space.
This quasi-particle picture of anyons is similar to that 
of two-dimensional anyons. However, the exotic 
statistics is not restricted to be the fractional 
statistics, but becomes a generalization 
from fractions to space-dependent functions. 
We further show that the scalar field is  
interpreted as the chiral mode of the system with chiral anomaly. \\

\acknowledgments

We thank F. T. Smith, C.L. Li, 
and I. Kulikov for their helpful discussions. 
The material is based upon research  supported in part by the
M\&H  Ferst Foundation, and by the NSF,  Grant No. PHY9211036.\\

\newpage
\begin{center}
\appendix 
\section*{A}
\end{center}
After a integral in part in Eq. (25),
we have
\begin{equation}
A_{0}(x)=\frac{e}{\lambda}\epsilon_{ij}\int d^{2}{\bf y}~
\frac{\partial}{\partial y^{i}}G({\bf x}-{\bf y})j^{j}(y).
\end{equation}
Using $\frac{\partial}{\partial y^{i}}G({\bf x}-{\bf y})=-
\frac{\partial}{\partial x^{i}}G({\bf x}-{\bf y})$,
and Eq. (8), we get 
\begin{equation}
A_{0}(x)=-\frac{e}{2\pi\lambda}\int d^{2}y~
\frac{\partial}{\partial x^{j}}\theta({\bf x}-{\bf y})j^{j}(y).
\end{equation}
Since $\frac{\partial}{\partial y^{i}}\theta({\bf x}-{\bf y})=-
\frac{\partial}{\partial x^{i}}\theta({\bf x}-{\bf y})$,
a further integral in part and using continuity equation of
the current, we get
\begin{equation}
A_{0}(x)=\frac{e}{2\pi\lambda}\int d^{2}
\theta({\bf x}-{\bf y})\partial_{0}j^{0}(y),
\end{equation}
which is exactly Eq. (26).\\

\begin{center}
\appendix 
\section*{B}
\end{center}
After a integral in part in Eq. (54), and using Eq. (48),
we get
\begin{equation}
A_{0}({\bf x})=-\int d^{3}{\bf x}^{\prime} \epsilon^{ijk}
\frac{\partial}{\partial x'^{j}}M_{k}({\bf x}-
{\bf x}')~ 
\left[\frac{D_{i}(x')}{\chi(x')}\right].
\end{equation}
After a further integral in part in  above equation 
and using Eq. (45), we get
\begin{equation}
A_{0}({\bf x})=-\int d^{3}{\bf x}^{\prime} \epsilon^{ijk}
M_{k}({\bf x}-{\bf x}')~ 
\frac{\partial}{\partial x'^{j}}\partial_{0} A_{i}(x').
\end{equation}
Using again Eq. (45) the expression of $C^{i}(x)$,
we directly obtain Eq. (55).

\begin{center}
\appendix 
\section*{C}
\end{center}
It follows from Eq. (45) that 
\begin{equation}
0=\frac{\partial}{\partial x^{i}}\left[
\frac{C^{i}(x)}{\chi(x)}\right]=
\left[\frac{\partial}{\partial x^{i}}C^{i}(x)\right]/\chi(x)-
\frac{\partial}{\partial x^{i}}\chi(x) C^{i}(x)/\chi(x)^{2}. 
\end{equation}
Using Eq. (43), we get
\begin{equation}
\frac{\partial\chi(x)}{\partial x_{i}}
\left[\frac{ C_{i}(x)}{\chi( x)}, 
~~\psi(x')\right]_{t=t'}=\frac{4\pi^{2}}{e}
\left[\rho(x),~\psi(x')\right]_{t=t'},
\end{equation}
which is easily reduced to be Eq. (60)
by adopting the anticommutation relation of 
fermion field:
\begin{equation}
\left\{\psi^{\dagger}(x),~\psi(x')\right\}_{t=t'}=
\delta({\bf x}-{\bf x}').
\end{equation}

\begin{figure}
\caption{ (a) shows that a closed loop $l$ encircles
a scalar monopole at $O$, $x$-axis is 
the branch cut. (b) shows  a closed loop that winds the 
the scalar monopole at $O$ $n$ circles}
\label{reduced}
\end{figure}

\begin{figure}
\caption{ (a) The quasi-particle picture of the anyons:
a fermion (bigger $dot$) at ${\bf x}$ is surrounded by 
infinite number 
scalar monopoles (smaller $dot$) in the whole space.
(b) When two anyons at ${\bf x}$ and ${\bf x'}$ exchange
their positions, each of them must pass through a monopole
of others.}
\label{reduced}
\end{figure}

\begin{figure}
\caption{(a) indicates the regions:  $R_{1}$ for $z>0$, and 
 $R_{2}$ for $z<0$; For $z=0$, i.e., $x-y$ section, 
 $R_{1}$ and  $R_{2}$ are shown in (b).}
\label{reduced}
\end{figure}


\begin{thebibliography}{s2}
\bibitem{comment1}
The connection between topology and statistics has been
approached for example in Ref. [19] in the Skyrme model,
where the topology is global.
\bibitem{monopole}
It should be clarified that the 
 statistics of a system with magnetic monopoles 
as the background topology does 
not mean the statistics of magnetic monopole.
\end{thebibliography}
\end{document}